\documentclass{IEEEtran}
\usepackage{cite}
\usepackage{amsmath,amssymb,amsfonts}
\usepackage{algorithmic}
\usepackage{graphicx}
\usepackage{textcomp}
\usepackage{algorithmic}
\usepackage{algorithm}
\usepackage{array}
\usepackage[caption=false,font=normalsize,labelfont=sf,textfont=sf]{subfig}
\usepackage{stfloats}
\usepackage{url}
\usepackage{verbatim}
\usepackage{enumerate}
\usepackage{tikz}
\usepackage{bm}
\usepackage{threeparttable,booktabs}
\usepackage{fancyhdr} 
\usepackage{hyperref}
\definecolor{lime}{HTML}{A6CE39}
\DeclareRobustCommand{\orcidicon}{
\begin{tikzpicture}
\draw[lime, fill=lime] (0,0)
circle[radius=0.16]
node[white]{{\fontfamily{qag}\selectfont \tiny \.{I}D}}; 
\end{tikzpicture}
\hspace{-2mm}
}
\foreach \x in {A, ..., Z}{%
\expandafter\xdef\csname orcid\x\endcsname{\noexpand\href{https://orcid.org/\csname orcidauthor\x\endcsname}{\noexpand\orcidicon}}
}

\fancypagestyle{firstpageheader}{
    \fancyhf{} 
    \fancyhead[C]{\textbf{\fontsize{8}{10}\selectfont This paper has been accepted by IEEE Transactions on Antennas and Propagation, DOI: \href{https://doi.org/10.1109/TAP.2024.3450298}{10.1109/TAP.2024.3450298}}} 
}

\def\BibTeX{{\rm B\kern-.05em{\sc i\kern-.025em b}\kern-.08em
    T\kern-.1667em\lower.7ex\hbox{E}\kern-.125emX}}
\begin{document}
\title{Scattering-Based Characteristic Mode Theory for Structures in Arbitrary Background: Computation, Benchmarks, and Applications
}

\author{Chenbo Shi\hspace{-1.5mm}\orcidA{}\hspace{-1mm}, Jin Pan\hspace{-1.5mm}\orcidB{}\hspace{-1mm}, Xin Gu, Shichen Liang and Le Zuo
\thanks{Manuscript received Dec. 29, 2023; revised Jul. 9, 2024; accepted in Aug. 6, 2024. This work was supported by the Aeronautical Science Fund under Grant ASFC-20220005080001. (\textit{Corresponding author: Jin Pan.})}
\thanks{Chenbo Shi, Jin Pan, Xin Gu and Shichen Liang are with the School of Electronic Science and Engineering, University of Electronic Science and Technology of China, Chengdu 611731 China  (e-mail: chenbo\_shi@163.com; panjin@uestc.edu.cn; xin\_gu04@163.com; lscstu001@163.com).}
\thanks{Le Zuo is with The 29th Research Institute of China Electronics Technology Group Corporation (e-mail: zorro1204@163.com)}
}

\markboth{}%
{Shell \MakeLowercase{\textit{et al.}}: A Sample Article Using IEEEtran.cls for IEEE Journals}

\maketitle
\thispagestyle{firstpageheader}

\begin{abstract}
  This paper presents a novel approach for computing substructure characteristic modes. This method leverages electromagnetic scattering matrices and spherical wave expansion to directly decompose electromagnetic fields. Unlike conventional methods that rely on the impedance matrix generated by the method of moments (MoM), our technique simplifies the problem into a small-scale ordinary eigenvalue problem, improving numerical dynamics and computational efficiency. We have developed analytical substructure characteristic mode solutions for a scenario involving two spheres, which can serve as benchmarks for evaluating other numerical solvers. A key advantage of our method is its independence from specific MoM frameworks, allowing for the use of various numerical methods. This flexibility paves the way for substructure characteristic mode decomposition to become a universal frequency-domain technique.
\end{abstract}


\section{Introduction}
\IEEEPARstart{C}{haracteristic} mode theory (CMT) reveals the inherent electromagnetic (EM) properties of structures, providing a comprehensive range of responses to external stimuli. Its applications in antenna design are diverse and grows rapidly \cite{ref_review1,ref_review2,ref_review3}. CMT was originally proposed by Garbacz as a universal method for decomposing scattering fields, based on scattering matrices and spherical waves expansion \cite{ref_TCM_Doctor,ref_TCM_Origin1}. However, due to challenges in implementing this method in mainstream numerical approaches, Harrington's method of moments (MoM) is often used instead \cite{ref_TCM_Origin2}. This substitution can alter the fundamental formula of characteristic modes, depending on the integral equations and equivalence strategies chosen \cite{ref_diff_MoM1,ref_diff_MoM2}, sometimes leading to ambiguous or even spurious modes \cite{ref_cross_validating}. A recent study \cite{ref_CM_unify} has made significant contributions by developing methods to calculate the scattering matrix, aligning the characteristic mode solving with Garbacz's original definition, and resolving issues related to the MoM. This approach also improves numerical stability, reduces computational costs for complex problems, and broadens the application of CM decomposition to include any frequency-domain solver that can generate scattering matrices.

Despite these advancements made by standard CMT, it has not resolved all inherent issues. Notably, the radiation properties derived from CMT for certain antenna types---such as dielectric resonant antennas, microstrip antennas, Yagi-Uda antennas, handheld \& wearable antennas, and platform-mounted antennas---are often at odds with prior knowledge \cite{ref_sTCM_Huang2,ref_sTCM_Microstrip,ref_comp_TCM1}. This discrepancy arises because the radiating part of these antennas is only a fraction of their entire structure, with the rest acting mainly as a background environment. For instance, in wearable antennas, the actual radiators are the devices on the body, not the body tissues. Standard CMT has overlooked such critical information, leading to deviations in the predicted mode properties from the actual antenna designs. To address these limitations, a variant called ``substructure characteristic theory'' has been proposed \cite{ref_sTCM_Concept,ref_sTCM_Review,ref_sTCM_Huang1}. We suggest renaming it to ``characteristic mode theory amidst certain background (termed BCMT)'' to more accurately reflect its principles and application scope.

In the approach for calculating the BCM, structures are first categorized into key (or controllable) and background (or loaded) structures. The Green's function for the background structures is used to establish the integral equation for the key structure area, which is then solved using the MoM. Such an extracted characteristic mode (i.e., BCM) considers the background's priori constraints and accurately reflects the antenna's intrinsic properties within a specific environment \cite{ref_sTCM_Huang2,ref_sTCM_APP1,ref_sTCM_APP2,ref_sTCM_APP3}. The Green's function of the background structure can be analytic, numerical, or matrix-based \cite{ref_Matrix_Green1,ref_Matrix_Green2,ref_Matrix_Green3}. The last allows for a generalized eigenvalue decomposition of the Schur component of the total impedance matrix obtained by the MoM, facilitating the extraction of BCM without needing to predict the background's Green's function (which is also the original method of BCMT \cite{ref_sTCM_Concept}). However, this method introduces ambiguity regarding the classification (key or background) of basis functions shared by the interface between key and background structures. This ambiguity necessitates the use of contact-region modeling (CRM) techniques \cite{ref_CRM}, inheriting all its drawbacks such as poor matrix condition and large size, which limits the BCMT's applicability to some extent. Moreover, like standard CMT, BCMT still faces challenges from its dependence on the MoM \cite{ref_CM_unify}.

Inspired by the achievements in \cite{ref_CM_unify}, this article aims to redefine the basic formulation of BCMT using the scattering matrix \cite{ref_S_mat1,ref_S_mat2,ref_S_mat3}, thereby eliminating its dependence on the MoM and enhancing computational efficiency. Assuming the background structure is lossless, we propose a novel formulation with concise modifications based on \cite{ref_TCM_Doctor} and \cite{ref_CM_unify}. This method effectively extracts the characteristic modes of structures amidst various finite-scale backgrounds. We prove that our formula is algebraically identical to the classical method based on the matrix Green's function, and show that Garbacz's original work is a special case of our approach when applied in a free space background. Our method also enables the analytical solving of spherical structures' BCM, which can serve as benchmarks for validating various BCM solvers. The relevant code is publicly available on \cite{ref_code}.

The descriptions in this article directly link to field quantities, offering significant advantages. It supports far-field based eigentrace tracking \cite{ref_tracking1,ref_tracking2,ref_unify2}, which improves performance over modal current correlation methods without increasing computational burden. Additionally, this approach transforms large-scale generalized eigenvalue equations into smaller-scale standard eigenvalue problems, enhancing numerical stability and reducing computational time. Due to the unique properties of the scattering matrix, our BCMT formulation remains consistent across different strategies for solving scattering matrices. Therefore, any method capable of solving the scattering matrix \cite{ref_S_solver,ref_unify2,ref_calc_T0,ref_calc_T1,ref_calc_T2,ref_calc_T3,ref_calc_T4,ref_Project_matrix} can be used to compute BCM, allowing BCMT to become a universal frequency technique, not limited to MoM. We also provide practical examples of antenna radiation and scattering analysis, demonstrating the considerable promise of BCMT in electromagnetic engineering.

\section{Scattering \& Transition Matrices and their Computations}
\label{SecII}

This section provides a concise overview of the transition and scattering matrices commonly used in scattering theory \cite{ref_CM_unify,ref_S_mat1,ref_S_mat2,ref_S_mat3,ref_scattering_theory}, introducing the notation to be used throughout this article. We consider a penetrable scatterer, $\Omega$, which is excited by incident fields $\vec{E}^{\mathrm{inc}}$ and $\vec{H}^{\mathrm{inc}}$, resulting in equivalent electromagnetic currents $\vec{J}_e$ and $\vec{J}_m$, and producing a scattered field $\vec{E}^s$ and $\vec{H}^s$. It is assumed that the source of the incident fields is located outside the smallest spherical surface $\Sigma$ that circumscribes $\Omega$. Additionally, the electromagnetic field is assumed to be time-harmonic, with the time convention $e^{\mathrm{j}\omega t}$. Under these conditions, the incident field can be described by a superposition of spherical wavefunctions:
\begin{equation}
  \label{eq1}
  \begin{cases}
    \displaystyle\vec{E}^{\mathrm{inc}}=k\sqrt{\eta}\sum_{\alpha}{\mathrm{a}_{\alpha}\vec{u}_{\alpha}^{\left( 1 \right)}\left( k\vec{r} \right)}\\
    \displaystyle\vec{H}^{\mathrm{inc}}=\mathrm{j}\frac{k}{\sqrt{\eta}}\sum_{\alpha}{\mathrm{a}_{\bar{\alpha}}\vec{u}_{\alpha}^{\left( 1 \right)}\left( k\vec{r} \right)}\\
  \end{cases}
\end{equation}
and outside of $\Sigma $, the scattering field can be expressed as
\begin{equation}
  \label{eq2}
  \begin{cases}
    \displaystyle\vec{E}^s=k\sqrt{\eta}\sum_{\alpha}{\mathrm{f}_{\alpha}\vec{u}_{\alpha}^{\left( 4 \right)}\left( k\vec{r} \right)}\\
    \displaystyle\vec{H}^s=\mathrm{j}\frac{k}{\sqrt{\eta}}\sum_{\alpha}{\mathrm{f}_{\bar{\alpha}}\vec{u}_{\alpha}^{\left( 4 \right)}\left( k\vec{r} \right)}\\
  \end{cases}
\end{equation}
where $\vec{u}_{\alpha}^{\left( p \right)}$ denotes vector spherical waves (cf. \cite{ref_scattering_theory}, $\S$ 7). Here, the superscript $p$ indicates the type of wave: $p=1$ for a regular wave, and $p=3,4$ for incoming and outgoing waves, respectively. Placing a bar over the index $\alpha$ interchanges the roles of the TE and TM waves. $\mathrm{a}_\alpha$ and $\mathrm{f}_\alpha$ represent the spherical wave expansion coefficients for the incident and scattering fields, respectively. The symbol $\mathrm{j}$ denotes the imaginary unit.

Within the above notation, the transition matrix $\mathbf{T}$ of $\Omega$ creates a linkage between the incident vector and the scattering vector, reads
\begin{equation}
  \label{eq3}
  \mathbf{f}=\mathbf{Ta}.
\end{equation}

Given that the scattering field is generated by the radiation of $\vec{J}_e$ and $\vec{J}_m$, the expansion coefficients of the scattering field can also be directly expressed from these currents as \cite{ref_CM_unify,ref_Project_matrix}
\begin{equation}
  \label{eq4}
  \mathbf{f}=-\mathbf{P}^{\left( 1 \right)}\mathbf{I}^e+\bar{\mathbf{P}}^{\left( 1 \right)}\mathbf{I}_m=-\mathbf{PI}
\end{equation}
where $\mathbf{I}=\left[ \mathbf{I}^e,\mathbf{I}^m\right] ^t$ is the expansion coefficients vector of $\vec{J}_e$ and $\mathrm{j}\vec{J}_m$ under the real-valued basis functions $\{ \vec{\psi}_i\}$, viz,. 
\begin{equation}
  \label{eq5}
  \vec{J}_e=\sum_i{\mathrm{I}_{i}^{e}\vec{\psi}_i},\quad \mathrm{j}\vec{J}_m=\sum_i{\mathrm{I}_{i}^{m}\vec{\psi}_i}.
\end{equation}
The superscript $t$ represents the transpose of the matrix, and the projection matrix
\begin{equation}
  \label{eq6}
  \mathbf{P}=\left[ \begin{matrix}
    \mathbf{P}^{\left( 1 \right)}&		-\bar{\mathbf{P}}^{\left( 1 \right)}\\
  \end{matrix} \right] 
\end{equation}
maps the electromagnetic current flow onto the spherical wave expansion coefficients. The $\alpha i$-th element of $\mathbf{P}^{\left( 1 \right)}$ and $\bar{\mathbf{P}}^{\left( 1 \right)}$ read
\begin{equation}
  \label{eq7}
  \left[ \mathbf{P}^{\left( 1 \right)} \right] _{\alpha i}=k\sqrt{\eta}\left< \vec{u}_{\alpha}^{\left( 1 \right)},\vec{\psi}_i \right> _{\Omega},\ \left[ \bar{\mathbf{P}}^{\left( 1 \right)} \right] _{\alpha i}=\frac{k}{\sqrt{\eta}}\left< \vec{u}_{\bar{\alpha}}^{\left( 1 \right)},\vec{\psi}_i \right> _{\Omega}
\end{equation}
where the symmetric product is defined as
\begin{equation}
  \label{eq8}
  \left< \vec{A},\vec{B} \right> _{\Omega}=\int{\vec{A}\cdot \vec{B}\mathrm{d}\Omega}.
\end{equation}

Per our previous assumption that $ \Omega $ is penetrable, the electromagnetic current flow through $\Omega$ can be solved using MoM equation
\begin{equation}
  \label{eq9}
  \mathbf{ZI}=\mathbf{V}.
\end{equation}
Here, the impedance matrix $\mathbf{Z}$ is composed of $\mathbf{Z}^o + \mathbf{Z}^i$, where $\mathbf{Z}^o$ and $\mathbf{Z}^i$ correspond to the outer and inner problems derived from the symmetrized  Poggio-Miller-Chang-Harrington-Wu-Tsai (PMCHWT) integral equations. The voltage vector
\begin{equation}
  \label{eq10}
  \mathbf{V}=\left[ \begin{matrix}
    \left<\left\{ \vec{\psi}_i \right\} , \vec{E}^{\mathrm{inc}}\right> _{\Omega} \\
    \mathrm{j}\left<\left\{ \vec{\psi}_i \right\},\vec{H}^{\mathrm{inc}}\right> _{\Omega}\\
  \end{matrix} \right] =\mathbf{P}^t\mathbf{a}.
\end{equation}

By solving for the current vector $\mathbf{I}$ from \eqref{eq9} and subsequently left-multiplying by $-\mathbf{P}$, the relationship between the transition matrix $\mathbf{T}$ and the impedance matrix is established as
\begin{equation}
  \label{eq11}
  \mathbf{T} = -\mathbf{P} \mathbf{Z}^{-1} \mathbf{P}^t.
\end{equation}
Note that this formulation also holds under the assumption of either perfect electric conductor (PEC) or perfect magnetic conductor materials. In such cases, $\mathbf{Z}$ and $\mathbf{P}$ contain matrices that exclusively relate to the electric or magnetic current flows, respectively. For alternative integral equations \cite{ref_MoM1,ref_MoM2,ref_MoM3,ref_MoM4,ref_MoM5}, cf. Appendix C in \cite{ref_CM_unify} for the detail.

Decomposing regular waves into incoming and outgoing waves leads to the scattering matrix
\begin{equation}
  \label{eq12}
  \mathbf{S}=\mathbf{1}+2\mathbf{T}
\end{equation}
where $\mathbf{1}$ is identity matrix. For a lossless scatterer, the scattering matrix $\mathbf{S}$ is unitary, meaning
\begin{equation}
  \label{eq13}
  \mathbf{S}^{\dagger}\mathbf{S}=\mathbf{1}.
\end{equation}
Here, the superscript $^\dagger$ indicates the Hermitian adjoint (or conjugate transpose) of the matrix. Substituting from \eqref{eq12} into the unitarity condition and using the property of normal matrices (where $\mathbf{S}$ is normal), we have
\begin{equation}
  \label{eq14}
  \mathbf{T}^{\dagger}\mathbf{T}=\mathbf{TT}^{\dagger}=-\mathrm{Re}\left\{ \mathbf{T} \right\}.
\end{equation}

The scattering and transition operators are solely determined by the properties of the scatterer, such as its material, shape, and electrical scale, making them independent of the form of excitation. Therefore, they are convenient in studying the system's eigenstates. The physical interpretation of these matrices is that any arbitrary incident (incoming) function is linearly transformed into a scattering (outgoing) function with the system's perturbation \cite{ref_TCM_Doctor,ref_S_mat1,ref_S_mat2,ref_S_mat3}. Choosing different basis functions for expanding the electromagnetic field does not alter their physical representation but results in variations in the numerical values of these matrices. In this article, we employ spherical wave functions as the basis because the scattering field from a practical finite-scale structure always takes the form of a spherical wave away from the scatterer.

\section{Characteristic Modes Theory of Structures amidst Arbitrary Finite-Scale Background}
\label{SecIII}
\subsection{Fundamental Formulation Based on Scattering Matrices}
The standard CMT formula for a key structure $\Omega_k$ can be defined in several equivalent ways, one of which involves scattering matrix or transition matrix \cite{ref_TCM_Doctor,ref_TCM_Origin1,ref_CM_unify}
\begin{equation}
  \label{eq15}
  \mathbf{Sf}_n=s_n\mathbf{f}_n \  \mathrm{or}\  \mathbf{Tf}_n=t_n\mathbf{f}_n
\end{equation}
where $s_n=1+2t_n$. The eigenvector $\mathbf{f}_n$ represents characteristic response (outgoing or scattering) field, and the eigenvalues $s_n$ and $t_n$ relate to the power properties of the characteristic mode (cf. equation (18) in \cite{ref_CM_unify}). A mapping defined by $t_n=-\left( 1+\mathrm{j}\lambda _n \right) ^{-1}$ determines the characteristic value $\lambda _n$, which is crucial for identifying the inductive, capacitive, and resonant attributes of the mode \cite{ref_TCM_Origin2}. These characteristics are particularly significant in the analysis and design of antennas.

Despite its many conveniences, the standard formulation \eqref{eq15} is limited to extracting the modal response of a scatterer in free space. When a background structure $\Omega_b$ is present, \eqref{eq15} will yield the eigenresponse of the combined system $\Omega_k \cup \Omega_b$ in free space, rather than the eigenresponse of $\Omega_k$ with $\Omega_b$ as its background, i.e., BCM. Therefore, we recognize that \eqref{eq15} requires revision within such a scene.

\begin{figure}[!t]
  \centering
  \includegraphics[]{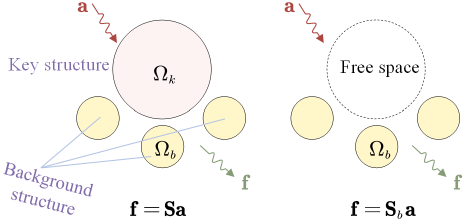}
  \caption{Configuration of the scattering matrices in \eqref{eq16}. To solve for $\mathbf{S}_b$, the key structure must be set to free space (i.e., removed).}
  \label{fEmbed_Smat}
\end{figure}

The modified formulation reads
\begin{equation}
  \label{eq16}
  \mathbf{SS}_{b}^{\dagger}\mathbf{f}_n=s_n\mathbf{f}_n
\end{equation}
where $\mathbf{S}$ is the scattering matrix for the entire system, encompassing both $\Omega_k$ and $\Omega_b$. $\mathbf{S}_b$ represents the scattering matrix of the background structure $\Omega_b$ alone, as illustrated in Fig. \ref{fEmbed_Smat}. This modification is crucial as it specifies which background is used to extract the characteristic modes. The equivalence between \eqref{eq16} and the impedance matrix-based definition of BCMT is further explored in Section \ref{SecIII_B}. 

In the sheer presence of free space, devoid of any scatterers, the natural transformation of incoming waves into outgoing waves designates the scattering matrix $\mathbf{S}_b=\mathbf{1}$ for free space. Our formulation, therefore, generalizes the computation of characteristic modes, while \eqref{eq15} is applicable as a special case in the absence of background structures. \eqref{eq16} also covers periodic structures\footnote{When analyzing periodic structures, the basis functions---spherical wave functions---should be replaced with Bloch harmonics \cite{ref_Floquet1,ref_Floquet2}}, but notice the deviation from the approach in \cite{ref_CM_Period}, where the product of the scattering matrices is interchanged to $\mathbf{S}_{b}^{\dagger}\mathbf{S}$. This interchange results in eigenvectors $\mathbf{f}^\prime_n$ that do not reside in the space spanned by outgoing wave functions, thus, $\mathbf{f}^\prime_n$ lack the interpretation as scattering fields. The algebraic connection between our work and \cite{ref_CM_Period} is captured by $\mathbf{f}_{n} = \mathcal{L}(\mathbf{S}_{b}\mathbf{f}^\prime_n)$, where $\mathcal{L}$ represents a linear transformation defined within the scattering function space. Given that $\mathbf{f}^\prime_n$ falls within the domain of the operator $\mathbf{S}_{b}$, it may be considered as a characteristic incoming field.

When the system is lossless, the unitary property
\begin{equation}
  \label{eq17}
  \mathbf{SS}_{b}^{\dagger}\cdot \left( \mathbf{SS}_{b}^{\dagger} \right) ^{\dagger}=\mathbf{1}
\end{equation}
indicates that the BCMs derived from \eqref{eq16} exhibit orthogonality:
\begin{equation}
  \label{eq18}
  \mathbf{f}_{m}^{\dagger}\mathbf{f}_n=\delta _{mn}
\end{equation}
provided that each $\mathbf{f}_n$ is normalized to $\left\| \mathbf{f}_n \right\|=1$. This orthogonality is equivalent to far-field orthogonality expressed as \cite{ref_TCM_Origin2}
\begin{equation}
  \label{eq19}
  \frac{1}{\eta}\underset{r\rightarrow \infty}{\lim}\oint_{4\pi}{\vec{E}_{m}^{*}\left( \hat{r} \right) \cdot \vec{E}_n\left( \hat{r} \right) \mathrm{d}S}=\delta _{mn}.
\end{equation}

To ensure the convergence of numerical calculations, we take an alternate description similar to the transition matrix
\begin{equation}
  \label{eq20}
  \hat{\mathbf{T}}\mathbf{f}_n=t_n\mathbf{f}_n
\end{equation}
where
\begin{equation}
  \label{eq21}
  \hat{\mathbf{T}}=\frac{\mathbf{SS}_{b}^{\dagger}-\mathbf{1}}{2}.
\end{equation}
However, it is important to note that $\hat{\mathbf{T}}$ is not the transition matrix for the scattering of $\Omega_k$ amidst the background $\Omega_b$, because $\mathbf{\hat f} = \hat{\mathbf{T}}\mathbf{a}$ does not equal to the scattering field of $\Omega_k$.

\subsection{Relationship with Impedance Matrix-based Formulation}
\label{SecIII_B}

The relationship between the scattering-based and impedance matrix-based characteristic mode formulations can be elucidated through equations \eqref{eq11} and \eqref{eq12}. To clarify this, we begin by partitioning the total impedance matrix $\mathbf{Z}$ and the projection matrix $\mathbf{P}$ into blocks, corresponding to the electromagnetic current basis functions associated with regions $\Omega_k$ and $\Omega_b$:
\begin{equation}
  \label{eq22}
  \mathbf{Z}=\left[ \begin{matrix}
    \mathbf{Z}_k&		\mathbf{Z}_{kb}\\
    \mathbf{Z}_{bk}&		\mathbf{Z}_b\\
  \end{matrix} \right] ,\quad \mathbf{P}=\left[ \begin{matrix}
    \mathbf{P}_k&		\mathbf{P}_b\\
  \end{matrix} \right].
\end{equation}
Then, an alternative expression of \eqref{eq11} in matrix blocks reads (cf. Appendix \ref{app_A})
\begin{equation}
  \label{eq23}
  \mathbf{T}=\tilde{\mathbf{T}}+\mathbf{T}_b
\end{equation}
with
\begin{equation}
  \label{eq24}
  \tilde{\mathbf{T}}=-\tilde{\mathbf{P}}\tilde{\mathbf{Z}}^{-1}\tilde{\mathbf{P}}^t,\quad \mathbf{T}_b=-\mathbf{P}_b\mathbf{Z}_{b}^{-1}\mathbf{P}_{b}^{t}
\end{equation}
where $\tilde{\mathbf{Z}}=\mathbf{Z}_k-\mathbf{Z}_{kb}\mathbf{Z}_{b}^{-1}\mathbf{Z}_{bk}$, $\tilde{\mathbf{P}}=\mathbf{P}_k-\mathbf{P} _b\mathbf{Z}_{b}^{-1}\mathbf{Z}_{bk}$. Here, we use $\mathbf{Z}_{kb}^{t}=\mathbf{Z}_{bk}$ as well as the symmetric properties of $\mathbf{Z}_k$ and $\mathbf{Z}_b$. Such defined matrix $\tilde{\mathbf{T}}$ functions similarly to a transition matrix for $\Omega_k$ amidst a background $\Omega_b$, which maps arbitrary incident field onto the scattering field in the presence of $\Omega_b$.

Right-multiplying $\mathbf{S}_b$ on both sides of \eqref{eq21} and utilizing \eqref{eq12} and \eqref{eq23} yields
\begin{equation}
  \label{eq25}
  \hat{\mathbf{T}}\mathbf{S}_b=\mathbf{T}-\mathbf{T}_b=\tilde{\mathbf{T}}
\end{equation}
then, with the equation \eqref{eq_bonds} in Appendix \ref{app_B}, we have
\begin{equation}
  \label{eq26}
  \hat{\mathbf{T}}=\tilde{\mathbf{T}}\mathbf{S}_{b}^{\dagger}=-\tilde{\mathbf{P}}\tilde{\mathbf{Z}}^{-1}\tilde{\mathbf{P}}^{\dagger}
\end{equation}
holds for lossless background.

Because the characteristic scattering field $\mathbf{f}_n$ arises from the radiation of characteristic electromagnetic current flow $\mathbf{I}_{k,n}$, with $\Omega _b$ serving as the background. By virtue of the scattering superposition principle, $\mathbf{f}_n$ is the sum of the direct radiation of $\mathbf{I}_{k,n}$ and the secondary radiation of $\mathbf{I}_{b,n}=-\mathbf{Z}_{b}^{-1}\mathbf{Z}_{bk}\mathbf{I}_{k,n}$ caused by background scatterer, reads 
\begin{equation}
  \label{eq27}
  \mathbf{f}_n=-\left( \mathbf{P}_k\mathbf{I}_{k,n}+\mathbf{P}_b\mathbf{I}_{b,n} \right) =-\tilde{\mathbf{P}}\mathbf{I}_{k,n}.
\end{equation}
Then, the eigenvalue equation of \eqref{eq26}, i.e., \eqref{eq20}, is equivalent to the impedance matrix-based generalized eigenvalue equation 
\begin{equation}
  \label{eq28}
  \tilde{\mathbf{Z}}\mathbf{I}_{k,n}=\left( 1+\mathrm{j}\lambda _n \right) \tilde{\mathbf{R}}\mathbf{I}_{k,n}
\end{equation}
with the radiation weighting operator $\tilde{\mathbf{R}}=\tilde{\mathbf{P}}^{\dagger}\tilde{\mathbf{P}}$, which can be shown by left multiplying $-t_n\tilde{\mathbf{P}}\tilde{\mathbf{Z}}^{-1}$ to \eqref{eq28}. For PEC's EFIE, $\tilde{\mathbf{R}}=\mathrm{Re}\left\{ \tilde{\mathbf{Z}} \right\}$; and for symmetrized PMCHWT of the penetrable structure, $\tilde{\mathbf{R}}=\mathrm{Re}\left\{ \mathbf{Z}_{k}^{o}-\mathbf{Z}_{kb}\mathbf{Z}_{b}^{-1}\mathbf{Z}_{bk} \right\} $, where we use $\mathbf{P}_{k}^{\dagger} \mathbf{P}_k=\mathrm{Re}\left\{ \mathbf{Z}_{k}^{o} \right\} $ \cite{ref_CM_unify}. These formulations are consistent with results in \cite{ref_sTCM_Concept} and \cite{ref_sTCM_Huang1,ref_sTCM_Huang2}. With $\mathbf{I}_{k,n}$ and $\tilde{\mathbf{R}}$, the orthogonality \eqref{eq18} of the characteristic modes can also be stated as \cite{ref_Orthogonality_BCMT,ref_Orthogonality_CMT}
\begin{equation}
  \label{eq29}
  \mathbf{I}_{k,m}^{\dagger}\tilde{\mathbf{R}}\mathbf{I}_{k,n}=\delta _{mn}.
\end{equation}

Solving for $\mathbf{I}_{k,n}$ from \eqref{eq28} results in
\begin{equation}
  \label{eq30}
  \mathbf{I}_{k,n}=\left( 1+\mathrm{j}\lambda _n \right) \tilde{\mathbf{Z}}^{-1}\tilde{\mathbf{R}}\mathbf{I}_{k,n}=t_{n}^{-1}\tilde{\mathbf{Z}}^{-1}\tilde{\mathbf{P}}^{\dagger}\mathbf{f}_n
\end{equation}
which can be used to reconstruct characteristic current from $\mathbf{f}_n$, obviating the need to re-solve the generalized eigenvalue equation \eqref{eq28}. Since $\tilde{\mathbf{Z}}^{-1}$ is already utilized in computing the total transition matrix $\mathbf{T}$ of the system, as shown in \eqref{eq23} and \eqref{eq24}, the application of \eqref{eq30} imposes only a minimal additional computational burden.

\subsection{Expansion Weighting of Characteristic Mode}

\begin{figure}[!t]
  \centering
  \includegraphics[]{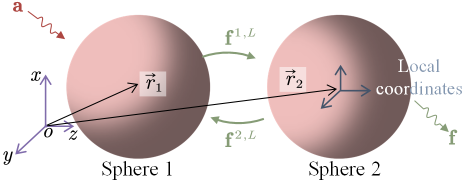}
  \caption{Schematic of dual-sphere scattering. $\vec r_1$ and $\vec r_2$ are their centers.}
  \label{ftwo_sph}
\end{figure}

Characteristic expansion weightings are utilized to quantify the extent and manner in which characteristic modes are excited. We assume that the external incident fields $\vec{E}^{\mathrm{inc}}$ and $\vec{H}^{\mathrm{inc}}$ excite both $\Omega _k$ and $\Omega _b$. Subsequently, the scattering response of $\Omega _b$ acts as a secondary source of excitation for the key structure $\Omega_k$. In the language of MoM, the direct excitation vectors for $\Omega _k$ and $\Omega _b$ are denoted as $\mathbf{V}_k$ and $\mathbf{V}_b$, respectively. Consequently, the total excitation vector for $\Omega _k$ is given by
\begin{equation}
  \label{eq31}
  \tilde{\mathbf{V}}=\mathbf{V}_k-\mathbf{Z}_{kb}\mathbf{Z}_{b}^{-1}\mathbf{V}_b
\end{equation}
where the term $-\mathbf{Z}_{kb} \mathbf{Z}_b^{-1} \mathbf{V}_b$ represents the secondary impaction of $\Omega_b$ on $\Omega_k$ through the coupled impedance matrices. Due to the equation \eqref{eq10}, \eqref{eq32} gives an alternative expression of \eqref{eq31}:
\begin{equation}
  \label{eq32}
  \tilde{\mathbf{V}}=\mathbf{P}_{k}^{t}\mathbf{a}-\mathbf{Z}_{kb}\mathbf{Z}_{b}^{-1}\mathbf{P}_{b}^{t}\mathbf{a}=\tilde{\mathbf{P}}^t\mathbf{a}.
\end{equation}

The weighting coefficient $w_n$ can be determined by starting with the equation $\tilde{\mathbf{V}}=\tilde{\mathbf{Z}}\tilde{\mathbf{I}}$, expanding $\tilde{\mathbf{I}}=\sum_n{w_n\mathbf{I}_{k,n}}$,  and then left multiplying by $\mathbf{I}_{k,m}^\dagger$ yields
\begin{equation}
  \label{eq33}
  w_n=\frac{\mathbf{I}_{k,n}^{\dagger}\tilde{\mathbf{V}}}{1+\mathrm{j}\lambda _n}=-t_n\mathbf{I}_{k,n}^{\dagger}\tilde{\mathbf{V}}
\end{equation}
due to \eqref{eq28} and the orthogonality condition \eqref{eq29}. Incorporating \eqref{eq32} and utilizing equation \eqref{eq_bonds} from Appendix \ref{app_B} results in an equivalent expression that involves only the expanding vector of spherical waves:
\begin{equation}
  \label{eq34}
  w_n=-\frac{\mathbf{f}_{n}^{\dagger}\mathbf{S}_b\mathbf{a}}{1+\mathrm{j}\lambda _n}=t_n\mathbf{f}_{n}^{\dagger}\mathbf{S}_b\mathbf{a}.
\end{equation}
Thus, the scattering field of $\Omega_k$ with $\Omega_b$ acting as the background can be expressed as
\begin{equation}
  \label{eq35}
  \mathbf{f}=\sum_n{w_n\mathbf{f}_n}.
\end{equation}

It is important to emphasize that our derivation eliminates the impractical assumption of zero excitation on the background structure $\Omega_b$ \cite{ref_sTCM_Concept}, common in real-world scenarios like scattering problems involving plane wave excitation. Furthermore, our formulation enables the use of BCMT in analyzing antennas excited by field coupling, treating the excitation structure as a source-containing background.

\section{Benchmark of Characteristic Mode Theory amidst Certain Background}
\label{SecIV}

The formulation \eqref{eq16} enables the analytical computation of BCMs for special structures, such as sphere and circular disk, which can serve as benchmarks for validating the performance of various BCM solvers. Since the transition and scattering matrices for a single spherical structure are well-understood ($\S$ 8 of \cite{ref_scattering_theory}), we examine a system comprising two spheres, as illustrated in Fig. \ref{ftwo_sph}. We assume that one sphere, denoted as $q$ (where $q=1\ \mathrm{or}\ 2$), acts as the background structure, while the other sphere, denoted as $\bar{q}$ (where $\bar{q}=2\ \mathrm{or}\ 1$), acts as the key structure. By leveraging transforming techniques for spherical wave functions, it becomes possible to analytically solve for the total scattering matrix of the system \cite{ref_scattering_theory,ref_translation1,ref_translation2}, allowing for the subsequent eigenvalue decomposition using \eqref{eq16}.

\subsection{Transition Matrix for Two-sphere System}

The transition matrix of the system inscribes the total scattering response $\vec{E}^s$ to external excitation $\vec{E}^{\mathrm{inc}}$ (with expansion vector $\mathbf{a}$) in free space, such a scattering field contains the common contribution from sphere $q=1\ \mathrm{and}\ 2$. The outside scattering field resulted from sphere $q$ has the form
\begin{equation}
  \vec{E}^{sq}\left( \vec{r}-\vec{r}_q \right) =\sum_{\alpha}{\mathrm{f}_{\alpha}^{q,L}\vec{u}_{\alpha}^{\left( 4 \right)}\left( \vec{r}-\vec{r}_q \right)}
\end{equation}
like \eqref{eq2}, where $\mathrm{f}^{q,L}_\alpha$ denotes the expansion coefficient of $\vec{E}^{sq}$ in the local coordinate system originates at sphere center $\vec{r}_q$. Throughout this article, the superscript ``$q,L$'' indicates using local coordinate system of sphere $q$.

\begin{figure*}[!t]
  \centering
  \subfloat[]{\includegraphics[]{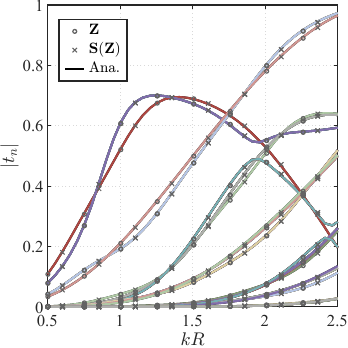}%
  \label{fMS_eg1}}
  \hfil
  \subfloat[]{\includegraphics[]{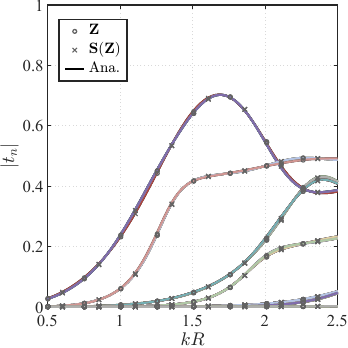}%
  \label{fMS_eg2}}
  \hfil
  \subfloat[]{\includegraphics[]{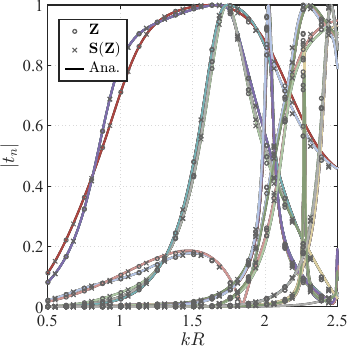}%
  \label{fMS_eg3}}
  \caption{Modal significance, where the line labeled $\mathbf{Z}$ is obtained by impedance matrix-based method; the line labeled $\mathbf{S(Z)}$ is obtained using \eqref{eq11}, \eqref{eq12} and \eqref{eq20}; the line labeled ``Ana.'' is obtained by the analytical methods. (a) involves two PEC spheres, (b) involves two dielectric spheres with one being lossy, and (c) involves one PEC sphere and one layered sphere.}
\label{fMS_eg}
\vspace{-0.15in}
\end{figure*}

Although the total scattering field in space is the summation of $\vec{E}^{sq}$ and $\vec{E}^{s\bar{q}}$, this does not imply that $\mathbf{f} = \mathbf{f}^{q,L} + \mathbf{f}^{\bar{q},L}$ due to the different coordinate systems involved. To reconcile these differences, it is necessary to transform these vectors into the global coordinate system (as discussed in \cite{ref_scattering_theory} and detailed in Appendix \ref{app_C}), viz.,
\begin{equation}
  \label{eq37}
  \mathbf{f}^q=\bm{\mathcal{R}}\left( k\vec{r}_{q} \right) \mathbf{f}^{q,L}
\end{equation}
where the elements of the matrix $\bm{\mathcal{R}} \left( \cdot \right) $ are given in Appendix \ref{app_C}. Then, 
\begin{equation}
  \mathbf{f}=\mathbf{f}^q+\mathbf{f}^{\bar{q}}.
\end{equation}

The properties $\bm{\mathcal{R}} ^{-1}( \vec{d}) =\bm{\mathcal{R}}( -\vec{d}) =\bm{\mathcal{R}} ^t( \vec{d}) $ also give the translation of the vector $\mathbf{f}^q$ into the local coordinate system of sphere $q$
\begin{equation}
  \label{eq39}
  \mathbf{f}^{q,L}=\bm{\mathcal{R}} ^t\left( k\vec{r}_{q} \right) \mathbf{f}^q.
\end{equation}
Analogously, the transformation of incident vector $\mathbf{a}$ into the local coordinate systems reads
\begin{equation}
  \mathbf{a}^{q,L}=\bm{\mathcal{R}} ^t\left( k\vec{r}_{q} \right) \mathbf{a}.
\end{equation}

The scattering of sphere $q$ arises from both the direct excitation by $\mathbf{a}^{q,L}$ and the secondary excitation from the scattering vector $\mathbf{f}^{\bar{q},L}$. However, due to the use of different basis functions, the total excitation is not simply the sum $\mathbf{a}^{q,L}+\mathbf{f}^{\bar{q},L} $. Instead, by employing the translation equation
\begin{equation}
  \mathbf{a}^{\bar{q}\rightarrow q,L}=\bm{\mathcal{Y}}^t \left( k\vec{d}_{q\bar{q}} \right) \mathbf{f}^{\bar{q},L}
\end{equation}
where $\vec{d}_{q\bar{q}}=\vec r_q-\vec r_{\bar q}$, we address this discrepancy. Consequently, the total incident waves on sphere $q$ are represented as $\mathbf{a}^{q,L}+\mathbf{a}^{\bar{q}\rightarrow q,L}$. Then, the transition matrix $\mathbf{T}^{q,L}$ of sphere $q$ dictates that
\begin{equation}
  \label{eq42}
  \mathbf{f}^{q,L}=\mathbf{T}^{q,L}\left[ \bm{\mathcal{R}}_{q}^{t}\mathbf{a}+\bm{\mathcal{Y}}^t_{q\bar{q}}\mathbf{f}^{\bar{q},L} \right].
\end{equation}
For convenience, we have denoted the arguments of the matrices $\bm{\mathcal{R}}$ and $\bm{\mathcal{Y}}$ with subscripts.

Substituting \eqref{eq42} into \eqref{eq37} and utilizing \eqref{eq39} yields
\begin{equation}
  \mathbf{f}^q=\bm{\mathcal{R}} _{q}\mathbf{T}^{q,L}\left[ \bm{\mathcal{R}} _{q}^{t}\mathbf{a}+\bm{\mathcal{Y}}^t_{q\bar{q}}\bm{\mathcal{R}}_{\bar{q}}^{t}\mathbf{f}^{\bar{q}} \right]
\end{equation}
let $q$ iterate through $\{1,2\}$, we can solve for
\begin{equation}
  \begin{split}
  \mathbf{f}^q=&\bm{\mathcal{R}} _{q}\left( \mathbf{1}-\mathbf{T}^{q,L}\bm{\mathcal{Y}}^t_{q\bar{q}}\mathbf{T}^{\bar{q},L}\bm{\mathcal{Y}}^t_{\bar{q}q} \right) ^{-1}\\
  &\times\mathbf{T}^{q,L}\left( \bm{\mathcal{R}}_{q}^{t}+\bm{\mathcal{Y}}^t_{q\bar{q}}\mathbf{T}^{\bar{q},L}\bm{\mathcal{R}}_{\bar{q}}^{t} \right) \mathbf{a}.
  \end{split}
\end{equation}
Therefore, the total transition matrix for the system reads
\begin{equation}
  \label{eq45}
  \mathbf{T} = \sum_{q=1,2} 
  \begin{aligned}
    \bm{\mathcal{R}}_{q} &\left ( \mathbf{1}-\mathbf{T}^{q,L} \bm{\mathcal{Y}}^t_{q\bar{q}} \mathbf{T}^{\bar{q},L} \bm{\mathcal{Y}}^t_{\bar{q}q} \right )^{-1} \\
    &\times \mathbf{T}^{q,L} \left( \bm{\mathcal{R}}_{q}^{t} + \bm{\mathcal{Y}}^t_{q\bar{q}} \mathbf{T}^{\bar{q},L} \bm{\mathcal{R}}_{\bar{q}}^{t} \right).
  \end{aligned}
 \end{equation}
\subsection{Transition Matrix for Sphere Alone}

Now consider a scenario where only sphere $q$ exists. Its transition matrix $\mathbf{T}^{q,L}$, relative to the local coordinate system, is a diagonal matrix directly derived from the boundary conditions imposed on the sphere. For instance, for a PEC sphere, the diagonal elements of $\mathbf{T}^{q,L}$ read
\begin{equation}
  \left\{ \begin{matrix}
    t_{1l,1l}^{q,L}=-\psi _l\left( kR \right) /\xi _l\left( kR \right)\\
    t_{2l,2l}^{q,L}=-\psi _{l}^{\prime}\left( kR \right) /\xi _{l}^{\prime}\left( kR \right)\\
  \end{matrix} \right.
\end{equation}
analogous to the Mie scattering coefficients, where $\psi _l\left( z \right) =zj_l\left( z \right) ,\xi _l\left( z \right) =zh_{l}^{\left( 2 \right)}\left( z \right) $, $R$ is the radius of this sphere. Using the translation formulas, the transition matrix $\mathbf{T}^q$ of sphere $q$ in the global coordinate system reads
\begin{equation}
  \label{eq47}
  \mathbf{T}^q=\bm{\mathcal{R}} _{q}\mathbf{T}^{q,L}\bm{\mathcal{R}}_{q}^{t}.
\end{equation}
For other spherical structures used in this article, only the term $\mathbf{T}^{q,L}$ in \eqref{eq47} needs to be replaced, which is detailed in $\S$ 8 of \cite{ref_scattering_theory}.

Leveraging \eqref{eq12}, \eqref{eq45} and \eqref{eq47}, we obtain the total scattering matrix $\mathbf{S}=\mathbf{1}+2\mathbf{T}$, and the scattering matrix $\mathbf{S}_b=\mathbf{1}+2\mathbf{T}^q$ for the background. Then, substituting these matrices into \eqref{eq20} will yield the benchmark results of BCMs.

\subsection{Benchmark Cases}

\subsubsection{PEC Spheres---Validating Numerical Dynamics}
\label{benchcase1}

\begin{figure}[!t]
  \centering
  \includegraphics[]{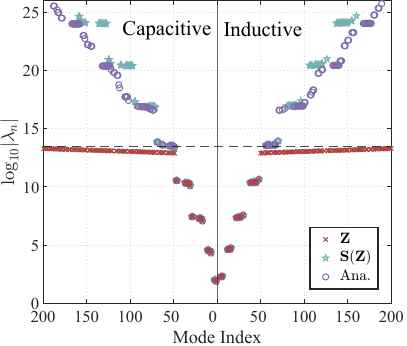}
  \caption{Characteristic values of BCMs for benchmark case Sec. \ref{benchcase1}.}
  \label{fCV_dynamic}
\end{figure}

This example involves two PEC spheres of equal radii, separated by a distance three times their radius. Fig. \ref{fMS_eg1} shows the modal significance (MS, i.e., $\left| t_n \right|$) curves for this setup, where results from three different methods align perfectly. Further analysis of the eigenvalue properties reveals that the impedance matrix-based method has a threshold with the increasing order of BCMs, beyond which accuracy is lost, as shown in Fig. \ref{fCV_dynamic}. However, the scattering-based method does not exhibit this issue, aligning with the analytic results. This phenomenon, known as numerical dynamics \cite{ref_CM_unify,ref_Project_matrix}, illustrates that the formulation proposed in this article behaves better numerical dynamics. 

\subsubsection{Dielectric Spheres---Validating Lossy Structures}
\label{benchcase2}

\begin{figure}[!t]
  \centering
  \subfloat[]{\includegraphics[]{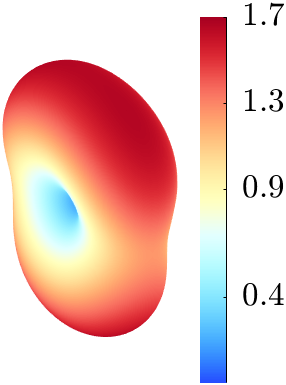}%
  \label{fbenchRad_J}}
  \hfil
  \subfloat[]{\includegraphics[]{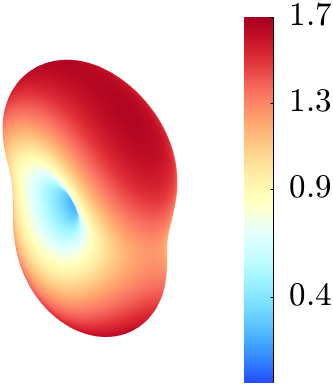}%
  \label{fbenchRad_T}}
  \hfil
  \subfloat[]{\includegraphics[]{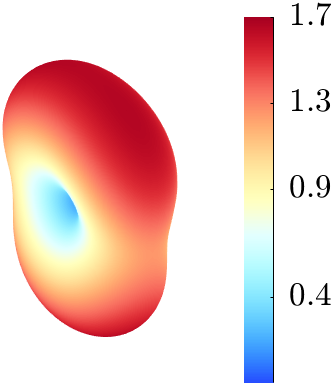}%
  \label{fbenchRad_A}}
  \caption{Characteristic far-field patterns for the dominant mode of benchmark case Sec. \ref{benchcase2} at $k2R=1$. (a) impedance matrix based method, (b) scattering-based method, (c) analytic method.}
\label{fbenchRad}
\end{figure}

In this example, we examine a lossy dielectric sphere with a radius of $0.75R$ and a relative permittivity of $\epsilon_r = 8 - 2\mathrm{j}$. This sphere is positioned against a background consisting of a dielectric sphere with a radius of $R$ and a relative permittivity of $\epsilon_r = 2$. The separation distance between the two spheres is three times their radius. Fig. \ref{fMS_eg2} displays the MS curves of the extracted BCMs, demonstrating perfect alignment of results obtained by different methods. Given the lossy nature of the structure, the modal significance $\left| t_n \right|$ are less than 1. This agreement is similarly observed when comparing their characteristic far-field patterns, as shown in Fig. \ref{fbenchRad}.

\subsubsection{Layered Spheres---Validating Eigentrace Tracking}
\label{benchcase3}
\begin{figure}[!t]
  \centering
  \subfloat[]{\includegraphics[]{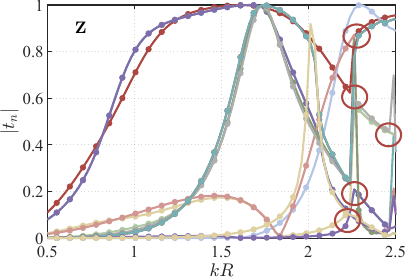}%
  \label{fMS_trackJ}}
  \vfil
  \subfloat[]{\includegraphics[]{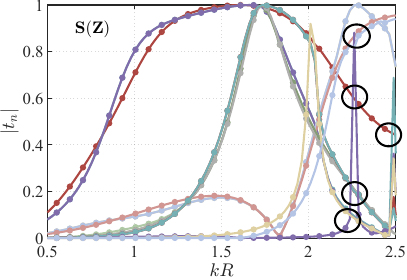}%
  \label{fMS_trackT}}
  \caption{Eigentrace tracking of benchmark case Sec. \ref{benchcase3}, the number of sampling points is 91. (a) is based on the correlation of characteristic current vectors. (b) is based on the correlation of characteristic scattering vectors. There were at least five tracking errors in (a), which were highlighted by red circles. }
\label{fMS_track}
\end{figure}

For the third example, we examine a PEC sphere with a radius $0.8R$, coated by a dielectric with a thickness of $\Phi = 0.2R$ and a relative permittivity of $\epsilon_r = 15$. This structure is positioned close to another PEC sphere with a radius $R$, and separated by a distance of $d = 3R$. The MS results are displayed in Fig. \ref{fMS_eg3}, where all methods exhibit consistent outcomes. However, the performance of eigentrace tracking differs. Fig. \ref{fMS_track} shows the trajectories of the tracked eigenvalues, clearly indicating that the scattering-based method outperforms the impedance matrix-based method.

Although both the impedance matrix-based and scattering-based approaches implement tracking procedures based on maximizing the correlation between the eigenvectors at successive sampling points ($x$ and $x+1$), their underlying principles differ significantly. Specifically, the impedance matrix-based method utilizes the following criterion:
\begin{equation}
  \underset{n}{\max}\left| \mathbf{I}_{k,m}^{\dagger}\left( x \right) \mathbf{I}_{k,n}\left( x+1 \right) \right|
\end{equation}
assuming each $\mathbf{I}_{k,m}$ is normalized such that $\mathbf{I}_{k,m}^{\dagger}\mathbf{I}_{k,m}=1$. Meanwhile, the scattering-based method applies:
\begin{equation}
  \label{eq49}
  \underset{n}{\max}\left| \mathbf{f}_{m}^{\dagger}\left( x \right) \mathbf{f}_n\left( x+1 \right) \right|.
\end{equation}
This approach, equivalent to far-field tracking \cite{ref_unify2}, tends to be more stable and performs better.

While far-field tracking could also be applied to the impedance matrix-based methods \cite{ref_tracking1,ref_tracking2}, it is notably time-consuming because it requires computing the radiated fields of all eigencurrents over a sphere and integrating them. In contrast, the computation for \eqref{eq49} merely involves the product of two vectors, making the computational burden negligible. Although other types of correlation coefficients, such as Pearson's coefficients \cite{ref_review3} and $| \mathbf{I}_{k,m}^{\dagger}\left( x \right) \tilde{\mathbf{R}}\mathbf {I}_{k,n}\left( x+1 \right)|$ \cite{ref_tracking_IRI}, can be used in the impedance matrix-based method, they do not significantly enhance performance relative to the additional costs in memory and time they incur.

\subsubsection{Computation Efficiency}

This section presents a comparative analysis of the computation times for the three examples using both the impedance matrix-based and scattering-based approaches. In the scattering-based approach, a total of 336 spherical waves were employed. Tables \ref{table_case1}, \ref{table_case2}, and \ref{table_case3} display the computational performance of these methods across the three examples.

For the first example, the problem scale is relatively small, making the performance of the two methods comparable. However, complexities arise in examples involving dielectric and layered materials. In such cases, the inclusion of magnetic currents doubles the number of unknowns, posing significant challenges for the impedance-based method in decomposing the large-scale matrix's generalized eigenvalues. Consequently, this method is limited to extracting only a few BCMs. In contrast, the scattering-based method exhibits superior performance, primarily because its major computational expense involves calculating the $\hat{\mathbf{T}}$ matrix, rather than performing eigenvalue decomposition, bacause the $\hat{\mathbf{T}}$ matrix is typically small scale.

\begin{table}[!t]
  \centering
  \caption{Evaluation of the computation time for benchmark case 1. 1071 of total 2145 RWGs are defined on key structure.}
  \begin{threeparttable}
  \begin{tabular}{ccc}
  \toprule
  Evaluated tasks & \begin{tabular}[c]{@{}c@{}}Scattering-based\\50/100/200 modes\end{tabular} & \begin{tabular}[c]{@{}c@{}}Impedance-based\\ 50/100/200 modes\end{tabular} \\ \midrule
  Calc. $\tilde{\mathbf{P}}$ matrix           & 3.22s           & -              \\ 
  Calc. $\tilde{\mathbf{Z}}$ matrix           & 2.88s           & 2.88s           \\ 
  Calc. $\hat{\mathbf{T}}$ matrix           & 0.08s           & -              \\ 
  Extract eigenvalues      & 0.67/0.68/0.72s & 3.19/4.15/4.21s \\ 
  Reconstruct currents     & 0.01/0.01/0.02s & -              \\ 
  Total time               & 6.86/6.87/6.92s & 6.07/7.03/7.09s \\ \bottomrule
  \end{tabular}
\end{threeparttable}
  \label{table_case1}
\end{table}

\begin{table}[!t]
  \centering
  \caption{Evaluation of the computation time for benchmark case 2. 4230 of total 11670 RWGs are defined on key structure.}
  \begin{threeparttable}
  \begin{tabular}{ccc}
  \toprule
  Evaluated tasks & \begin{tabular}[c]{@{}c@{}}Scattering-based\\ 50/100/200 modes\end{tabular} & \begin{tabular}[c]{@{}c@{}}Impedance-based\\ 50/100/200 modes\end{tabular} \\ \midrule
  Calc. $\tilde{\mathbf{P}}$ matrix          & 9.18s              & -                    \\ 
  Calc. $\tilde{\mathbf{Z}}$ matrix          & 81.96s             & 81.96s               \\ 
  Calc. $\hat{\mathbf{T}}$ matrix          & 1.82s              & -                    \\ 
  Extract eigenvalues     & 0.71/0.74/0.68s    & 12.5/34.25/57.05s    \\ 
  Reconstruct currents    & 0.21/0.25/0.33s    & -                    \\ 
  Total time              & 93.78/93.85/93.87s & 94.45/116.11/139.01s \\ \bottomrule
  \end{tabular}
  \end{threeparttable}
  \label{table_case2}
\end{table}

 \begin{table}[!t]
  \centering
  \caption{Evaluation of the computation time for benchmark case 3. 8253 of total 10368 RWGs are defined on key structure.}
  \begin{threeparttable}
  \begin{tabular}{ccc}
  \toprule
  Evaluated tasks & \begin{tabular}[c]{@{}c@{}}Scattering-based\\ 50/100/200 modes\end{tabular} & \begin{tabular}[c]{@{}c@{}}Impedance-based\\ 50/100/200 modes\end{tabular} \\ \midrule
  Calc. $\tilde{\mathbf{P}}$ matrix           & 11.72s                & -                     \\ 
  Calc. $\tilde{\mathbf{Z}}$ matrix           & 91.89s                & 91.89s                \\ 
  Calc. $\hat{\mathbf{T}}$ matrix           & 8.27s                 & -                     \\ 
  Extract eigenvalues      & 0.68/0.7/0.68s        & 47.49/109.94/176.04s  \\ 
  Reconstruct currents     & 0.1/0.15/0.22s        & -                     \\ 
  Total time               & 112.66/112.73/112.78s & 139.38/201.83/267.93s \\ \bottomrule
  \end{tabular}
  \end{threeparttable}
  \label{table_case3}
\end{table}

\section{Applications in Antenna Analysis}
\subsection{Expanding Scattering Field of Structure amidst Certain Background}

\begin{figure}[!t]
  \centering
  \subfloat[]{
    \begin{tikzpicture}
      \node (img1) {\includegraphics{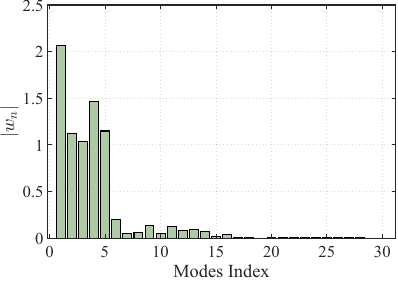}};
      \node at (img1.center)[shift={(1cm,0.9cm)}]{\includegraphics{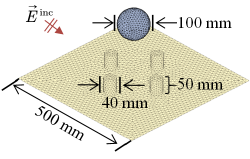}};
      \label{fRCS_alpha}
    \end{tikzpicture}
  }
  \vfil
  \subfloat[]{\includegraphics[]{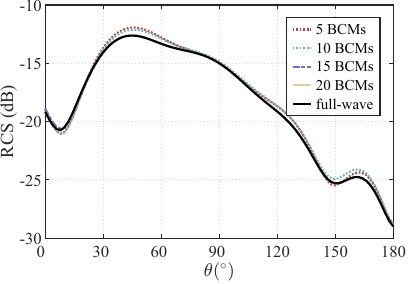}%
  \label{fRCS_diffnum}}
  \caption{(a) is the magnitude of characteristic expansion weightings for the first 30 BCMs. (b) is the superpositioned RCS using different numbers of BCMs.}
\end{figure}

\begin{figure}[!t]
  \centering
  \begin{tikzpicture}
    \node (img1) {\includegraphics{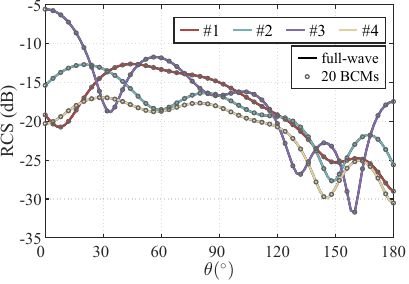}};
    \node at (img1.center)[shift={(-1.4cm,-0.57cm)}]{\includegraphics{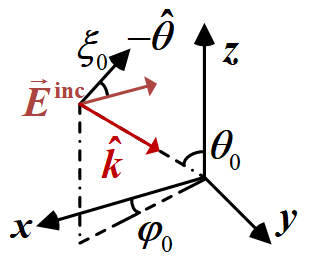}};
  \end{tikzpicture}
  \caption{RCS caused by different incident plane waves. The wave's directions $\left( \theta _0,\phi _0\right)$ and polarization angles $\xi_0$, i.e., $\left( \theta _0,\phi _0,\xi _0 \right)$ are $\left( 30^{\circ},0^{\circ},45^{\circ} \right)$ for \#1, $\left( 90^{\circ},30^{\circ},0^{\circ} \right)$ for \#2, $\left( 0^{\circ},0^{\circ},0^{\circ} \right)$ for \#3, and $\left( -120^{\circ},90^{\circ},0^{\circ} \right)$ for \#4.} 
  \label{fRCS_diffinc}
  \vspace{-0.15in}
\end{figure}

The BCMT enables the representation of the electromagnetic field scattered by a structure in non-free space using a limited number of modes. We illustrate this with a metallic sphere scatterer set against a complex metallic background, illuminated by a plane wave from the direction of $\left( \theta _0,\phi _0\right) =\left( 30^{\circ},0^{\circ}\right)$ with the polarization angle $\xi _0=45^{\circ}$ at 1 GHz (see Fig. \ref{fRCS_diffinc} for the definition of these angles). The spherical wave expansion coefficients for this plane wave (cf. $\S$ 7 of \cite{ref_scattering_theory}) determines the characteristic expansion weightings \eqref{eq34} for each BCM in Fig. \ref{fRCS_alpha}. Figure \ref{fRCS_diffnum} demonstrates how the RCS approaches the full-wave solution as the number of BCMs increases. At 20 modes, the maximum RCS error falls below 1\%.

The necessary number of BCMs typically depends on the electrical scale of the problem, not the specifics of the incident plane wave. For instance, Fig. \ref{fRCS_diffinc} demonstrates that employing a 20-mode expansion across various plane wave incidents---regardless of direction and polarization---yields results consistent with the full-wave solution. Thus, information from the first 20 modes suffices to address all scattering issues prompted by plane waves, suggesting that RCS design could be optimized by manipulating these 20 modes, particularly those with higher weighting coefficients.

\subsection{Analyzing Antenna's Radiation}

\begin{figure}[!t]
  \centering
  \includegraphics[]{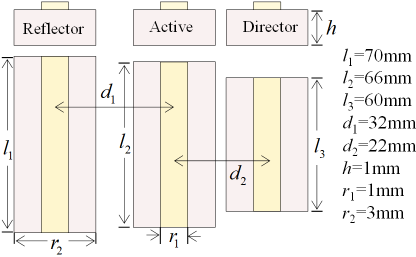}
  \caption{Prototype of the tri-element printed Yagi-Uda antenna.}
  \label{fYagi_Model}
\end{figure}

\begin{figure}[!t]
  \centering
  \includegraphics[]{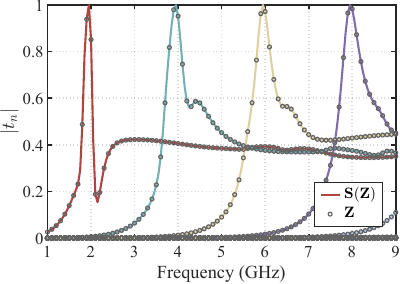}
  \caption{MS of the Yagi-Uda antenna.}
  \label{fMS_Yagi}
\end{figure}

\begin{figure}[!t]
  \centering
  \subfloat[]{\includegraphics[]{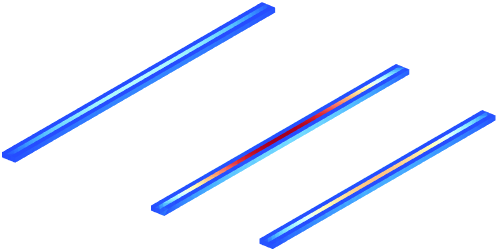}%
  \label{fCur_CM1}}
  \hfil
  \subfloat[]{\includegraphics[]{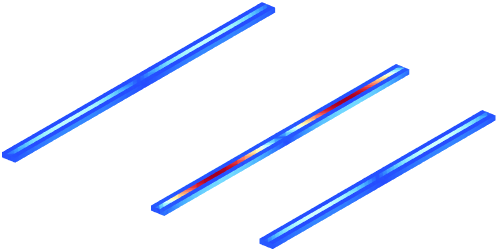}%
  \label{fCur_CM2}}
  \vfil
  \subfloat[]{\includegraphics[]{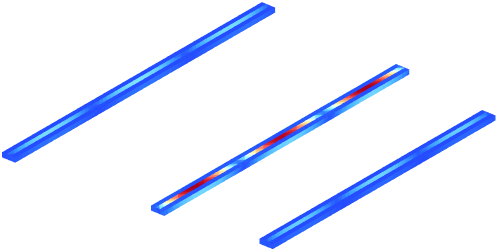}%
  \label{fCur_CM3}}
  \hfil
  \subfloat[]{\includegraphics[]{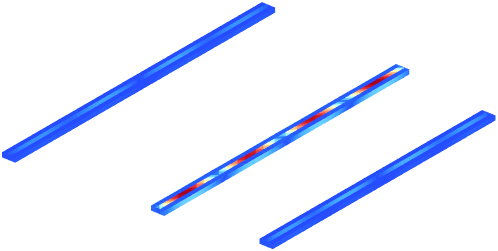}%
  \label{fCur_CM4}}
  \caption{Characteristic electric currents of the Yagi-Uda antenna. (a), (b), (c) and (d) show the respective currents of BCM1, BCM2, BCM3 and BCM4. In this figure, red areas represent higher magnitudes, while blue areas indicate lower magnitudes.}
  \label{fcur_CM}
\end{figure}

Analyzing characteristic modes for antennas can predict radiation properties effectively. We demonstrate this with a tri-element printed Yagi-Uda antenna, as depicted in Fig. \ref{fYagi_Model}. With priori knowledge, the central active dipole is identified as the key structure. Therefore, we performed a BCM decomposition using \eqref{eq16}, selecting $\mathbf{S}_b$ as the scattering matrix for the structure except the central dipole. The MS curves, shown in Fig. \ref{fMS_Yagi}, reveal four resonant modes between 1-9 GHz, with resonances at 1.95 GHz, 3.95 GHz, 6 GHz, and 7.95 GHz. The corresponding electric current distributions appear in Fig. \ref{fcur_CM}.

\begin{figure}[!t]
  \centering
  \includegraphics[]{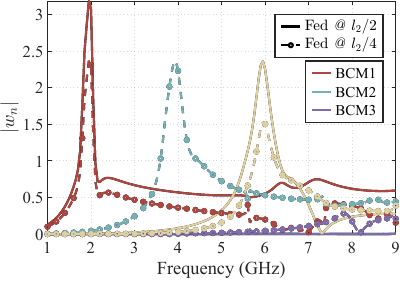}
  \caption{BCM expansion weightings of the Yagi-Uda antenna fed at different position. The voltage is 1 mV for the feeding port.}
  \label{falphaYagi}
\end{figure}

\begin{figure}[!t]
  \centering
  \includegraphics[]{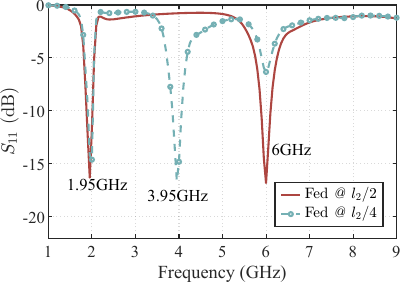}
  \caption{S11 of the Yagi-Uda antenna fed at different positions.}
  \label{fS11Yagi}
\end{figure}

\begin{figure}[!t]
  \centering
  \subfloat[]{\includegraphics[]{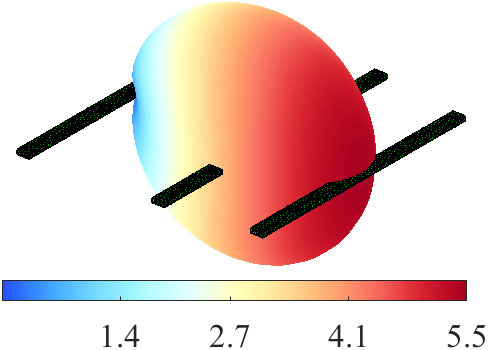}%
  \label{fRad_CM1}}
  \hfil
  \subfloat[]{\includegraphics[]{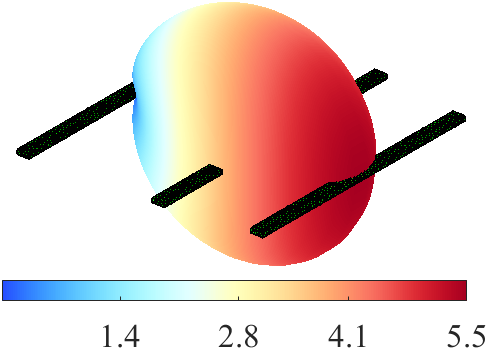}%
  \label{fRad_fed1}}
  \vfil
  \vspace{-0.12in}
  \subfloat[]{\includegraphics[]{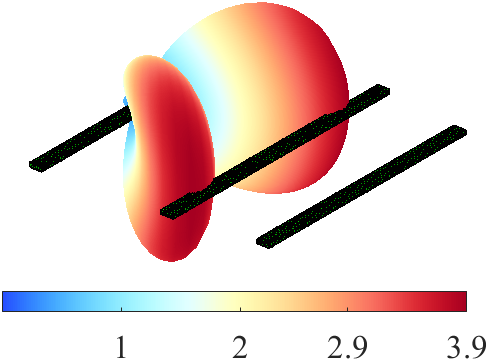}%
  \label{fRad_CM2}}
  \hfil
  \subfloat[]{\includegraphics[]{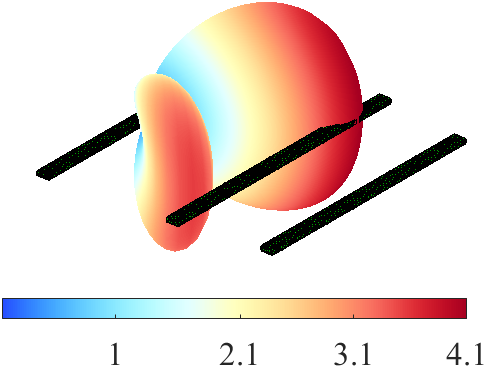}%
  \label{fRad_fed2}}
  \vfil
  \vspace{-0.12in}
  \subfloat[]{\includegraphics[]{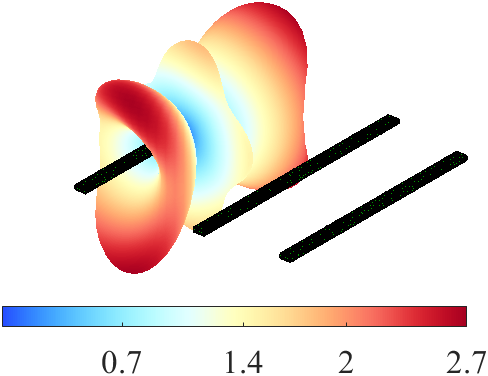}%
  \label{fRad_CM3}}
  \hfil
  \subfloat[]{\includegraphics[]{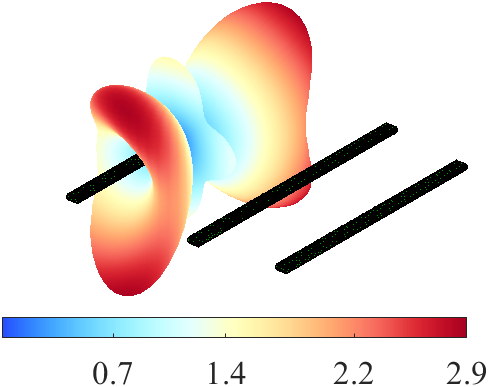}%
  \label{fRad_fed3}}
  \caption{Radiation patterns. The left panel represents BCM far-field patterns from BCM1 to BCM3, while the right panel represents the simulated radiation patterns using FEKO. (b) and (f) are fed at the center, while (d) is fed offset.}
  \vspace{-0.15in}
\end{figure}

Assuming an ideal voltage port feeds the antenna, we calculate BCM weightings using \eqref{eq33}. Because effective excitation requires maximizing energy coupling from the feeding port, i.e., ensuring the term $\mathbf{I}_{k,n}^{\dagger}\mathbf{V}$ in \eqref{eq33} as large as possible. By situating the voltage port at the active dipole---aligning with the reddest area of the characteristic currents---we infer that modes 1 and 3 can be excited effectively from the center. The resulting BCM weightings, shown in Fig. \ref{falphaYagi}, demonstrate that modes 1 and 3 dominate near their respective resonance frequencies. As BCM2 and BCM4 exhibit current nulls at the center, they are inhibited. As a result, we anticipate the S-parameters to peak near 1.95 GHz and 6 GHz, corresponding to BCM1 and BCM3, respectively, with similar radiation patterns. Figures \ref{fS11Yagi}, \ref{fRad_CM1}, \ref{fRad_fed1}, \ref{fRad_CM3}, and \ref{fRad_fed3} confirm these predictions.

To excite additional modes like BCM2, altering the feeding position is necessary. For instance, relocating the port to $1/4$ of the active dipole's length allows the excitation of BCM1 -- BCM3, as all exhibit significant current magnitudes. This modification is evident in Fig. \ref{falphaYagi} and \ref{fS11Yagi}, where the $S_{11}$ curve now displays three resonance peaks corresponding to respective BCM1 -- BCM3. However, exciting higher-order modes results in less pure mode components, causing deviations in the simulated radiation patterns from the BCM's far-field patterns.

This example underscores the utility of BCMT in foreseeing antenna radiation performance and its potential in practical antenna designs.

\section{Conclusion}

This paper presents a new formulation \eqref{eq16} of BCMT based on the scattering field (scattering matrix), which is an extension of Garbacz's original work. Our formulation has been proven fully equivalent to the impedance matrix-based formulation \eqref{eq28} in lossless backgrounds, but offers superior numerical stability, enhanced numerical dynamics, and reduced computational effort. Unlike \eqref{eq28}, our method's reliance on the unique properties of the scattering matrix ensures consistent performance across various computational strategies without the needing to alter the core formulation. Particularly for homogeneous and isotropic spherical structures, we have analytically solved their BCMs, providing definitive results that benchmark the efficacy of different BCMT solvers.

Thanks to the clear interpretation of \eqref{eq16}, we have resolved ambiguities in determining whether interface unknowns in layered structures should be categorized as key or background, eliminating the need for the CRM technique. Additionally, our framework could support the use of nonconformal domain decomposition techniques, beneficial for tackling special problems characterized by multiscale features and local refinements. Since any method capable of solving scattering problems can be used to solve scattering matrices, including widely-used commercial software like HFSS, CST, FEKO and COMSOL \cite{ref_S_solver}, therefore, our method facilitates the broader application of BCMT as a generalized frequency-domain method.

Beyond theoretical and computational advancements, we have present case studies on radiation and scattering analysis. These demonstrate BCMT's potential to foresee structural radiation performance, positioning it as a promising tool in antenna design.

\begin{appendices}
\section{Transition Matrix Calculated using Blocks of Impedance Matrix}
\label{app_A}
The representation of the $\mathbf{T}$ matrix using the blocks of impedance matrix can be obtained by first finding a blocked representation of $\mathbf{B}=\mathbf{Z}^{-1}\mathbf{P}^t$. For this purpose, we consider the matrix equation
\begin{equation}
  \mathbf{ZB}=\mathbf{P}^t.
\end{equation}
Bifurcating it into blocks, we have
\begin{equation}
  \left[ \begin{matrix}
    \mathbf{Z}_k&		\mathbf{Z}_{kb}\\
    \mathbf{Z}_{bk}&		\mathbf{Z}_b\\
  \end{matrix} \right] \left[ \begin{matrix}
    \mathbf{B}_k\\
    \mathbf{B}_b\\
  \end{matrix} \right] =\left[ \begin{matrix}
    \mathbf{P}_{k}^{t}\\
    \mathbf{P}_{b}^{t}\\
  \end{matrix} \right] 
\end{equation}
explicitly,
\begin{equation}
  \begin{cases}
    \mathbf{Z}_k\mathbf{B}_k+\mathbf{Z}_{kb}\mathbf{B}_b=\mathbf{P}_{k}^{t}\\
    \mathbf{Z}_{bk}\mathbf{B}_k+\mathbf{Z}_b\mathbf{B}_b=\mathbf{P}_{b}^{t}\\
  \end{cases}.
\end{equation}
The last row gives $\mathbf{B}_b=\mathbf{Z}_{b}^{-1}\mathbf{P}_{b}^{t}-\mathbf{Z}_{b}^{-1}\mathbf{Z}_{bk}\mathbf{B}_k$. By inserting it into the first row leads to
\begin{equation}
  \mathbf{B}_k=\tilde{\mathbf{Z}}^{-1}\tilde{\mathbf{P}}^t
\end{equation}
where $\tilde{\mathbf{Z}}=\mathbf{Z}_k-\mathbf{Z}_{kb}\mathbf{Z}_{b}^{-1}\mathbf{Z}_{bk}$, $\tilde{\mathbf{P}}=\mathbf{P}_k-\mathbf{P}_b\mathbf{Z}_{b}^{-1}\mathbf{Z}_{bk}$, and note that $\mathbf{Z}_{kb}^{t}=\mathbf{Z}_{bk}$, $\mathbf{Z}_k$ and $\mathbf{Z}_b$ are symmetric. Therefore,
\begin{equation}
  \mathbf{Z}^{-1}\mathbf{P}^t=\left[ \begin{matrix}
    \mathbf{B}_k\\
    \mathbf{B}_b\\
  \end{matrix} \right] =\left[ \begin{matrix}
    \tilde{\mathbf{Z}}^{-1}\tilde{\mathbf{P}}^t\\
    \mathbf{Z}_{b}^{-1}\mathbf{P}_{b}^{t}-\mathbf{Z}_{b}^{-1}\mathbf{Z}_{bk}\tilde{\mathbf{Z}}^{-1}\tilde{\mathbf{P}}^t\\
  \end{matrix} \right] .
\end{equation}
Thereby, an alternative expression of \eqref{eq11} reads
\begin{equation}
  \label{eqTblocks}
  \mathbf{T}=-\mathbf{P}\mathbf{B}=-\tilde{\mathbf{P}}\tilde{\mathbf{Z}}^{-1}\tilde{\mathbf{P}}^t+\mathbf{T}_b
\end{equation}
where $\mathbf{T}_b=-\mathbf{P}_b\mathbf{Z}_{b}^{-1}\mathbf{P}_{b}^{t}$.

\section{Critical Formula for Lossless Background Structure}
\label{app_B}
For lossless background structure $\Omega_b$, the properties \eqref{eq14} of $\mathbf{T}_b$ leads to
\begin{equation}
  \label{eq56}
  \mathbf{Z}_{b}^{-1}\mathbf{P}_{b}^{t}\mathbf{P}_b\mathbf{Z}_{b}^{-*} = \mathrm{Re}\left\{ \mathbf{Z}_{b}^{-1} \right\}
\end{equation}
then, the critical relation
\begin{equation}
  \label{eq_bonds}
  \tilde{\mathbf{P}}^{\dagger}=\tilde{\mathbf{P}}^t\left( 1+2\mathbf{T}_{b}^{\dagger} \right) =\tilde{\mathbf{P}}^t\mathbf{S}_{b}^{\dagger}
\end{equation}
holds. 

\noindent
\textbf{\textit{Proof:}} we first rewrite the term
\begin{equation}
  \label{eq58}
  \begin{split}
    &\mathrm{Re}\left\{ \mathbf{Z}_{kb}\mathbf{Z}_{b}^{-1} \right\}\\
    &\quad=\frac{1}{2}\left\{ \mathbf{Z}_{kb}\mathbf{Z}_{b}^{-1}+\left( \mathbf{Z}_{kb}\mathbf{Z}_{b}^{-1} \right) ^{*} \right\}
    =\frac{1}{2}\left\{ \mathbf{Z}_{kb}\mathbf{Z}_{b}^{-1}+\mathbf{Z}_{kb}^{*}\mathbf{Z}_{b}^{-*} \right\}\\
    &\quad=\frac{1}{2}\left\{ \mathbf{Z}_{kb}+\mathbf{Z}_{kb}^{*} \right\} \mathbf{Z}_{b}^{-*}-\frac{1}{2}\mathbf{Z}_{kb}\left\{ \mathbf{Z}_{b}^{-1}+\mathbf{Z}_{b}^{-*} \right\} +\mathbf{Z}_{kb}\mathbf{Z}_{b}^{-1}\\
    &\quad=\mathrm{Re}\left\{ \mathbf{Z}_{kb} \right\} \mathbf{Z}_{b}^{-*}-\mathbf{Z}_{kb}\mathrm{Re}\left\{ \mathbf{Z}_{b}^{-1} \right\} +\mathbf{Z}_{kb}\mathbf{Z}_{b}^{-1}.
  \end{split}
\end{equation}
Because $\mathrm{Re}\left\{ \tilde{\mathbf{P}}^t \right\} =\mathbf{P}_{k}^{t}-\mathrm{Re}\left\{ \mathbf{Z}_{kb}\mathbf{Z}_{b}^{-1} \right\} \mathbf{P}_{b}^{t}$, leveraging \eqref{eq58}, using \eqref{eq56} and $\mathrm{Re}\left\{ \mathbf{Z}_{kb} \right\} =\mathbf{P}_{k}^{t}\mathbf{P}_b$, we can show that
\begin{equation}
  \begin{split}
    &\mathrm{Re}\left\{ \tilde{\mathbf{P}}^t \right\}=\mathbf{P}_{k}^{t}-\mathbf{Z}_{kb}\mathbf{Z}_{b}^{-1}\mathbf{P}_{b}^{t}\\
    &\quad-\mathbf{P}_{k}^{t}\mathbf{P}_b\mathbf{Z}_{b}^{-*}\mathbf{P}_{b}^{t} + \mathbf{Z}_{kb}\mathbf{Z}_{b}^{-1}\mathbf{P}_{b}^{t}\mathbf{P}_b\mathbf{Z}_{b}^{-*}\mathbf{P}_{b}^{t}\\
    &\quad=\left( \mathbf{P}_{k}^{t}-\mathbf{Z}_{kb}\mathbf{Z}_{b}^{-1}\mathbf{P}_{b}^{t} \right) +\left( \mathbf{P}_{k}^{t}-\mathbf{Z}_{kb}\mathbf{Z}_{b}^{-1}\mathbf{P}_{b}^{t} \right) \mathbf{P}_b\mathbf{Z}_{b}^{-*}\mathbf{P}_{b}^{t}\\
    &\quad=\tilde{\mathbf{P}}^t\left( \mathbf{1}+\mathbf{T}_{b}^{\dagger} \right) .
  \end{split}
\end{equation}
Then, $\tilde{\mathbf{P}}^{t}=2\mathrm{Re}\left\{ \tilde{\mathbf{P}}^t \right\} -\tilde{\mathbf{P}}^{\dagger}$ leads to \eqref{eq_bonds}.

\section{Translation Properties of Spherical Wave Functions}
\label{app_C}
Let the origin $O$ separate from the origin $O^\prime$ by a distance vector $\vec d$, i.e. $\vec r=\vec r^\prime+\vec d$. Then, the spherical waves with respect to origin $O^\prime$ can be translated to origin $O$ using the following properties \cite{ref_scattering_theory,ref_translation1,ref_translation2}:

\textit{Translation Property 1---}
\begin{equation}
  \vec{u}_{\alpha}^{(4)}\left( k\vec{r} \right) =\sum_{\alpha ^{\prime}}{\bm{\mathcal{R}} _{\alpha \alpha ^{\prime}}}(k\vec{d})\vec{u}_{\alpha ^{\prime}}^{(4)}\left( k\vec{r}^{\prime} \right), r^\prime>d.
\end{equation}

\textit{Translation Property 2---}
\begin{equation}
  \vec{u}_{\alpha}^{(1)}\left( k\vec{r} \right) =\sum_{\alpha ^{\prime}}{\bm{\mathcal{R}}_{\alpha \alpha ^{\prime}}}(k\vec{d})\vec{u}_{\alpha ^{\prime}}^{(1)}\left( k\vec{r}^{\prime} \right), \forall \vec d.
\end{equation}

\textit{Translation Property 3---}

\begin{equation}
  \vec{u}_{\alpha}^{(4)}\left( k\vec{r} \right) =\sum_{\alpha ^{\prime}}{\bm{\mathcal{Y}}_{\alpha \alpha ^{\prime}}}(k\vec{d})\vec{u}_{\alpha ^{\prime}}^{(1)}\left( k\vec{r}^{\prime} \right),r^\prime<d.
\end{equation}
Cf. Appendix F.7 in \cite{ref_scattering_theory} can find the elements of matrices $\bm{\mathcal{R}}(k\vec d)$ and $\bm{\mathcal{Y}}(k\vec d)$. Here we assume that $\vec d$ is along positive $z$ axis and provide a simplified version, specifically,
\begin{equation}
  \begin{array}{l}
    \bm{\mathcal{Y}} _{1\sigma ml,1\sigma ml^{\prime}}(k\vec{d})=c_m\cdot C_{l,l^{\prime},m}\\
    \bm{\mathcal{Y}} _{1\sigma ml,1\sigma ^{\prime}ml^{\prime}}(k\vec{d})=0,\sigma\ne\sigma^\prime\\
    \bm{\mathcal{Y}} _{1\sigma ml,2\sigma ^{\prime}ml^{\prime}}(k\vec{d})=(-1)^{\sigma+m}D_{l,l^{\prime},m},\sigma\ne\sigma^\prime\\
    \bm{\mathcal{Y}} _{1\sigma ml,2\sigma ml^{\prime}}(k\vec{d})=0\\
    \bm{\mathcal{Y}} _{2\sigma ml,\tau \sigma ^{\prime}ml^{\prime}}(k\vec{d})=\bm{\mathcal{Y}} _{1\sigma ml,\bar{\tau}\sigma ^{\prime}ml^{\prime}}(k\vec{d}),\tau =1,2\\
  \end{array}
\end{equation}
and $\bm{\mathcal{R}}(k\vec d)=\mathrm{Re}\left \{\bm{\mathcal{Y}}(k\vec d)\right \}$. Here, $c_m=(-1)^m+\delta_{m0}(-1)^\sigma$, and the coefficients $C_{l,l^\prime,m}$ and $D_{l,l^\prime,m}$ read
\begin{equation}
  \begin{aligned}
    C_{l,l^{\prime},m}=&\frac{\varepsilon_m}{4}\times \sum_{\lambda =\left| l-l^{\prime} \right|}^{l+l^{\prime}}(-1)^{\left( l^{\prime}-l+\lambda \right) /2}(2\lambda +1)\\
    &\times \sqrt{\frac{(2l+1)\left( 2l^{\prime}+1 \right)}{l(l+1)l^{\prime}\left( l^{\prime}+1 \right)}}\\
    &\times \left( \begin{matrix}
    l&		l^{\prime}&		\lambda\\
    0&		0&		0\\
  \end{matrix} \right) \left( \begin{matrix}
    l&		l^{\prime}&		\lambda\\
    m&		-m&		0\\
  \end{matrix} \right)\\
    &\times \left[ l(l+1)+l^{\prime}\left( l^{\prime}+1 \right) -\lambda (\lambda +1) \right]\\
    &\times h^{(2)}_{\lambda}(kd)\\
  \end{aligned}
\end{equation}
and 
\begin{equation}
  \begin{aligned}
    D_{l,l^{\prime},m}&=-m\times \sum_{\lambda =\left| l-l^{\prime} \right|}^{l+l^{\prime}}(-1)^{\left( l^{\prime}-l+\lambda \right) /2}(2\lambda +1)\\
    &\times \sqrt{\frac{(2l+1)\left( 2l^{\prime}+1 \right)}{l(l+1)l^{\prime}\left( l^{\prime}+1 \right)}}\\
    &\times \left( \begin{matrix}
    l&		l^{\prime}&		\lambda\\
    0&		0&		0\\
  \end{matrix} \right) \left( \begin{matrix}
    l&		l^{\prime}&		\lambda\\
    m&		-m&		0\\
  \end{matrix} \right)\\   
    &\times kd\times h^{(2)}_{\lambda}(kd)\\
  \end{aligned}
\end{equation}
where $\varepsilon_m=2-\delta_{m0}$, and $\left( \begin{matrix}
	\cdot&		\cdot&		\cdot\\
	\cdot&		\cdot&		\cdot\\
\end{matrix} \right) $ represents Wigner 3-j symbol \cite{ref_Wigner}.

\end{appendices}

\begin{IEEEbiography}[{\includegraphics[width=1in,height=1.25in,clip,keepaspectratio]{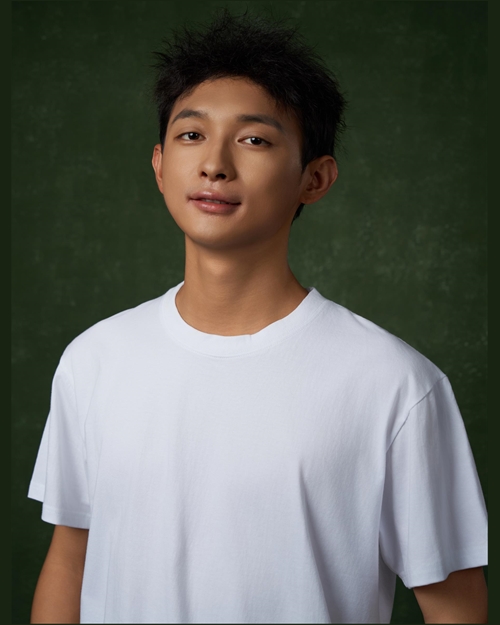}}]{Chenbo Shi}
Chenbo Shi was born in 2000 in China. He received his Bachelor's degree from the University of Electronic Science and Technology of China (UESTC) in 2022. He is currently pursuing his Ph.D. at the same institution. His research interests include electromagnetic theory, characteristic mode theory, and computational electromagnetics. 

Chenbo has been actively involved in several research projects and has contributed to publications in these areas. His work aims to advance the understanding and application of electromagnetic phenomena in various technological fields.
\end{IEEEbiography}

\begin{IEEEbiography}[{\includegraphics[width=1in,height=1.25in,clip,keepaspectratio]{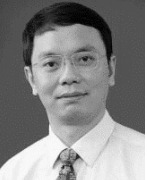}}]{Jin Pan}
received the B.S. degree in electronics and communication engineering from the Radio Engineering Department, Sichuan University, Chengdu, China, in 1983, and the M.S. and Ph.D. degrees in electromagnetic field and microwave technique from the University of Electronic Science and Technology of China (UESTC), Chengdu, in 1983 and 1986, respectively. 

From 2000 to 2001, he was a Visiting Scholar in electronics and communication engineering with the Radio Engineering Department, City University of Hong Kong, Hong Kong. 
He is currently a Full Professor with the School of Electronic Engineering, UESTC. 

His current research interests include electromagnetic theories and computations, antenna theories, and techniques, field and wave in inhomogeneous media, and microwave remote sensing theories and its applications. 
\end{IEEEbiography}

\begin{IEEEbiography}[{\includegraphics[width=1in,height=1.25in,clip,keepaspectratio]{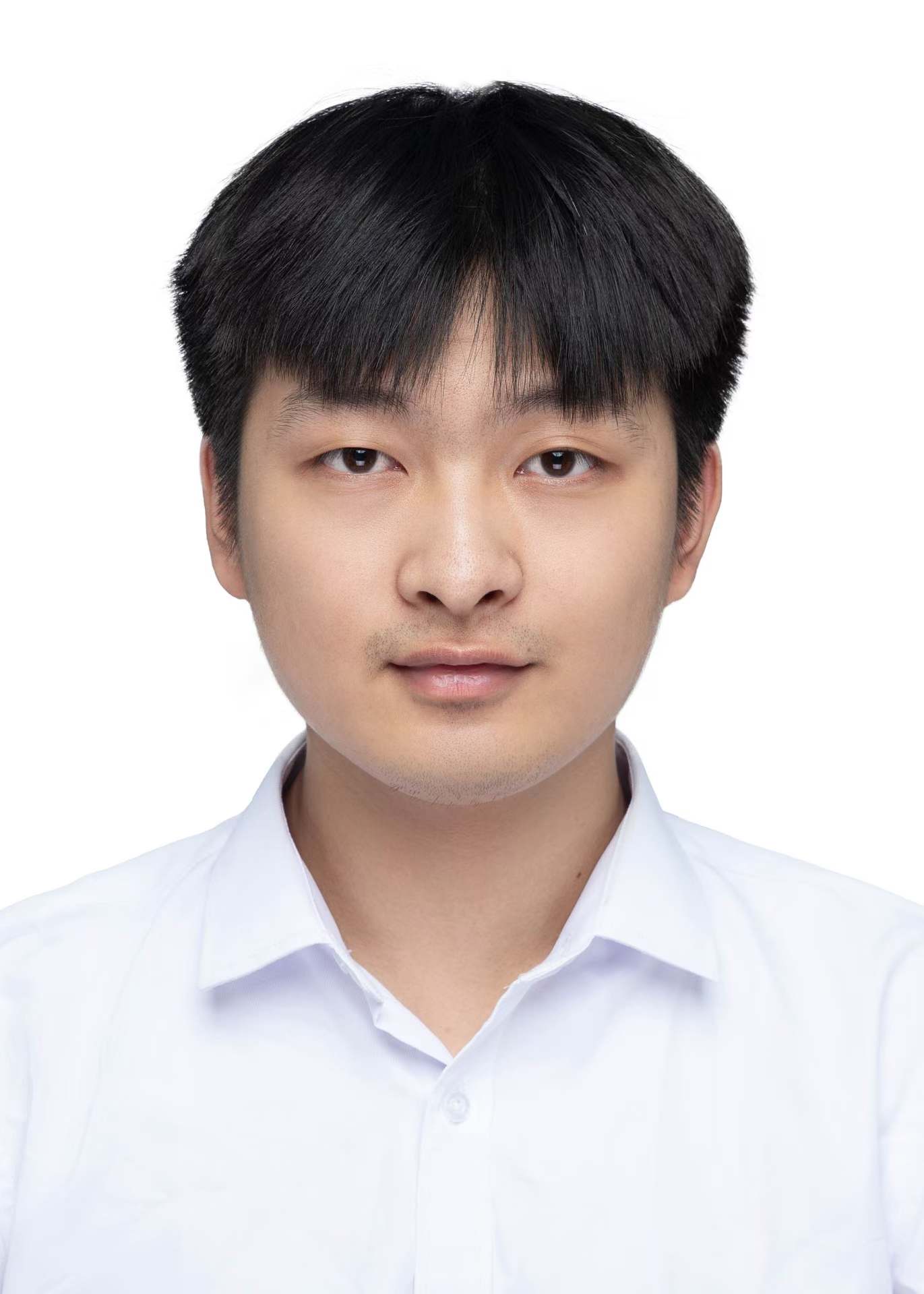}}]{Xin Gu}
received the B.E. degree from Chongqing University of Posts and Telecommunications (CQUPT), ChongQing, China, in 2022. He is currently pursuing the M.S. degree with the School of Electronic Science and Engineering, University of Electronic Science and Technology of China (UESTC), Chengdu, China.
	
His research interests include electromagnetic theory and electromagnetic measurement techniques
\end{IEEEbiography}

\begin{IEEEbiography}[{\includegraphics[width=1in,height=1.25in,clip,keepaspectratio]{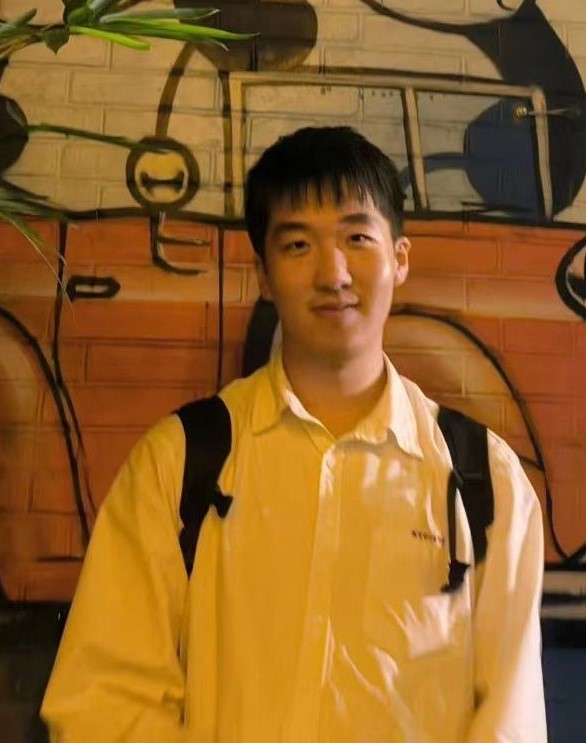}}]{Shichen Liang}
received the B.E. degree from Beijing University of Chemical Technology (BUCT), Beijing, China, in 2022. He is currently pursuing the M.S. degree with the School of Electronic Science and Engineering, University of Electronic Science and Technology of China (UESTC), Chengdu, China. 

His research interests include electromagnetic theory and electromagnetic measurement techniques.
\end{IEEEbiography}

\begin{IEEEbiography}[{\includegraphics[width=1in,height=1.25in,clip,keepaspectratio]{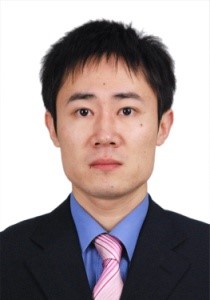}}]{Le Zuo}
received the B.Eng., M.Eng. and Ph.D. degrees in electromagnetic field and microwave techniques from the University of Electronic Science and Technology of China (UESTC), in 2004, 2007 and 2018, respectively. 

From 2017 to 2018, he was a Research Associate with the School of Electrical and Electronic Engineering, Nanyang Technological University, Singapore. He is currently a Research Fellow with the 29th Institute of the China Electronics Technology Group, Chengdu, China. His research interests include antenna theory and applications.
\end{IEEEbiography}

\end{document}